\documentclass[aps,pre,amsmath,amssymb,twocolumn,superscriptaddress,showpacs]{revtex4}     
\usepackage{graphicx}
\begin{document}

\title{Heterogeneous Nucleation in the Low Barrier Regime}
\author{Benjamin Scheifele}
\affiliation{Department of Physics, St. Francis Xavier University,
Antigonish, NS, B2G 2W5, Canada}

\author{Ivan Saika-Voivod}
\affiliation{Department of Physics and Physical Oceanography,
Memorial University of Newfoundland, St. John's, NL, A1B 3X7, Canada}

\author{Richard K. Bowles}
\affiliation{Department of Chemistry, University of Saskatchewan, Saskatoon, SK, 57N 5C9, Canada}

\author{Peter H. Poole}
\affiliation{Department of Physics, St. Francis Xavier University, Antigonish, NS, B2G 2W5, Canada}

\begin{abstract}
In simulations of the 2D Ising model, we examine heterogeneous nucleation induced by a small impurity consisting of a line of $l$ fixed spins.  As $l$ increases, we identify a limit of stability beyond which the metastable phase is not defined.  We evaluate the free energy barrier for nucleation of the stable phase, and show that contrary to expectation, the barrier does not vanish on approach to the limit of stability.  We also demonstrate that our values for the height of the barrier yield predictions for the nucleation time (from transition state theory) and the size of the critical cluster (from the nucleation theorem) that are in excellent agreement with direct measurements, even near the limit of stability.
\end{abstract}


\date{\today}
\maketitle

\section{Introduction}

The formalism of transition state theory (TST), as developed by Volmer and Weber~\cite{vol26}, Becker and D\"{o}ring~\cite{becker35}, Zeldovich~\cite{zeldo}, and Frenkel~\cite{frenkel46}, 
continues to be of fundamental importance for understanding phase transformations in a great variety of systems.  The assumptions upon which TST is based nominally restrict this approach to predicting nucleation rates for systems that are only mildly metastable and for which the nucleation barrier is large relative to $kT$, where $T$ is the temperature and $k$ is Boltzmann's constant. However, many interesting phase changes occur in the deeply metastable regime where the system is approaching a limit of stability and the free energy barrier to nucleation is expected to disappear~\cite{cahn,BZ,TO,parrinello06,vanish,S}. Some recent simulation studies~\cite{SPB,Maibaum08,reguera09} find that the predictions of TST remain surprisingly robust in this deeply metastable regime, but others suggest that TST breaks down~\cite{parrinello06,bagchi07,bagchi11}.  Understanding whether TST remains applicable, or how the formalism should be adapted, when the nucleation barrier becomes low remains an open question.

The presence of a heterogeneous interface in a metastable system can dramatically lower the nucleation barrier in phase transformations such as vapor condensation and crystallization~\cite{pablo, kelton}. Consequently, heterogeneous nucleation plays an important role in a variety of phenomena including atmospheric physics~\cite{CK,kulmala08,winkler08}, the use of templates to form complex structures~\cite{zak98,cacciuto04,cacciuto05} and protein crystallization~\cite{chayen06,S07}. The basic principles of heterogeneous nucleation involving macroscopic, bulk surfaces are relatively well established.  However, in many cases the heterogeneities are microscopic in size and there is considerable interest in understanding how particle size influences the nucleation mechanism and rate~\cite{oh01,winkler08,cacciuto04,cacciuto05,sear,kea10,rkb11}, especially as the barrier approaches $kT$.

Here we study heterogeneous nucleation in the two-dimensional ($2D$) Ising model to explore the nature of the nucleation barrier on approach to the limit of stability of a metastable phase.  We seek to clarify the definition of the barrier in this limit, and to test the degree to which theories (in particular TST, and also the nucleation theorem) are able to predict the behaviour observed directly in this regime.  The Ising system we examine was studied previously by Sear~\cite{sear}, who demonstrated that a small cluster of fixed ``impurity" spins increased the nucleation rate significantly relative to the homogeneous nucleation rate.  In the present study, we exploit the fact that by increasing the size of the impurity we can systematically lower the nucleation barrier and also bring the system to a limit of stability.  At the same time, as we will show, this simple model allows the heterogeneous nucleation barrier to be defined in a way that is free of significant approximations that affect the definition of the homogeneous nucleation barrier when the barrier height is low.  As a consequence, this model provides an excellent opportunity to compare 
the free energy barrier, critical cluster size, and nucleation rate
as predicted by theory, with values obtained by direct simulations.

\section{Methods}

Our results are based on Monte Carlo (MC) simulations of a 2D Ising model of a ferromagnet.   We employ a $L\times L$ square lattice with periodic boundary conditions, and choose $L=45$, the same system size studied in Ref.~\cite{sear}.  The energy of the system in spin configuration $c$ is given by,
\begin{equation}
E_c=-J\sum_{\langle i,j \rangle} s_i s_j + H\sum_{i=1}^N s_i
\label{ham}
\end{equation}
where $s_i=\pm 1$ is the spin value of site $i$, $J>0$ quantifies the ferromagnetic exchange interaction, $H$ is the value of the external magnetic field, and $N=L^2$ is the number of sites in the lattice.  The sum in the first term is taken over all nearest-neighbor pairs of spins.  We explore the configuration space of the system using Metropolis single-spin-flip MC dynamics, in which one Monte Carlo step (MCS) corresponds to $N$ spin-flip attempts, and where spins are chosen at random.  

In each of our runs, we initialize all free spins to $s_i=-1$, and equilibrate the system in the spin-down phase at $\beta J=0.65$ (i.e. 0.678 of the critical temperature) and $\beta H=-0.05$, where $\beta=1/kT$.  We then create a metastable state by instantaneously changing the sign of the magnetic field, so that $\beta H=+0.05$.  These choices of $T$ and $H$ are the same as those used in Ref.~\cite{sear}.  Under these conditions the spin-down phase is metastable, and the system persists in this phase until nucleation of the stable spin-up phase occurs.  

\section{Homogeneous Nucleation}

The aim of the present work is to study a system in which heterogeneous nucleation is the dominant process for transforming the metastable to the stable phase.  In this section, we quantify the homogeneous nucleation process that occurs when no impurity is present.  Doing so allows us to confirm that we are working under conditions where homogeneous nucleation can be neglected once we introduce an impurity into the system.  We emphasize that we are not attempting here to conduct a detailed examination of homogeneous nucleation in the Ising model.  There have been a number of recent and very thorough studies of homogeneous nucleation in the Ising model, to which we refer the interested reader~\cite{pan,brendel,Cai2010a,Cai2010b}.

We begin by evaluating the free energy barrier for homogeneous nucleation, following the same approach as used in Ref.~\cite{brendel,Cai2010a,Cai2010b}.
This method exploits the fact that when up-spin clusters occurring in a metastable down-spin phase are rare and do not interact, the free energy $g$ to form an up-spin cluster of size $m$ is well approximated by,
\begin{equation}
\beta g(m)=-\log\frac{{\cal N}(m)}{N}.
\label{homo}
\end{equation}
where ${\cal N}(m)$ is the average number of up-spin clusters of size $m$~\cite{thesis,reiss1999,auer2004}.

As we will see, the homogenous nucleation barrier in our case is large relative to $kT$.  As in Refs.~\cite{brendel,Cai2010a,Cai2010b}, we therefore use an umbrella sampling method to access the relatively rare configurations of the system that occur near the top of the nucleation barrier.  In this approach, a biasing potential $U_B=\kappa(M-M_0)^2$ is added to the system potential energy given in Eq.~\ref{ham}, where $M$ is the size of the largest cluster of up-spins in the system, and $M_0$ is a target value of $M$.  The effect of $U_B$ is to drive the system to sample configurations for which $M$ is close to $M_0$, over a range of $M$ that is controlled by the value of the parameter $\kappa$.  For a given value of $M_0$, we determine a segment of the ${\cal N}(m)$ curve from,
\begin{equation}
{\cal N}(m)=\bigl\langle {\cal N}_B(m)  \exp[\beta U_B(M,M_0) ] \bigl\rangle,
\label{use}
\end{equation}
where ${\cal N}_B(m)$ is the number of up-spin clusters of size $m$ for a system configuration sampled during the biased simulation.  In Eq.~\ref{use}, $\langle \cdots \rangle$ denotes an ensemble average computed during the biased simulation, and the exponential factor reweights the result to provide the estimate of ${\cal N}(m)$ that would be found from an unbiased simulation (i.e. one with $U_B=0$).

\begin{figure}
\bigskip
\centerline{\includegraphics[scale=0.35]{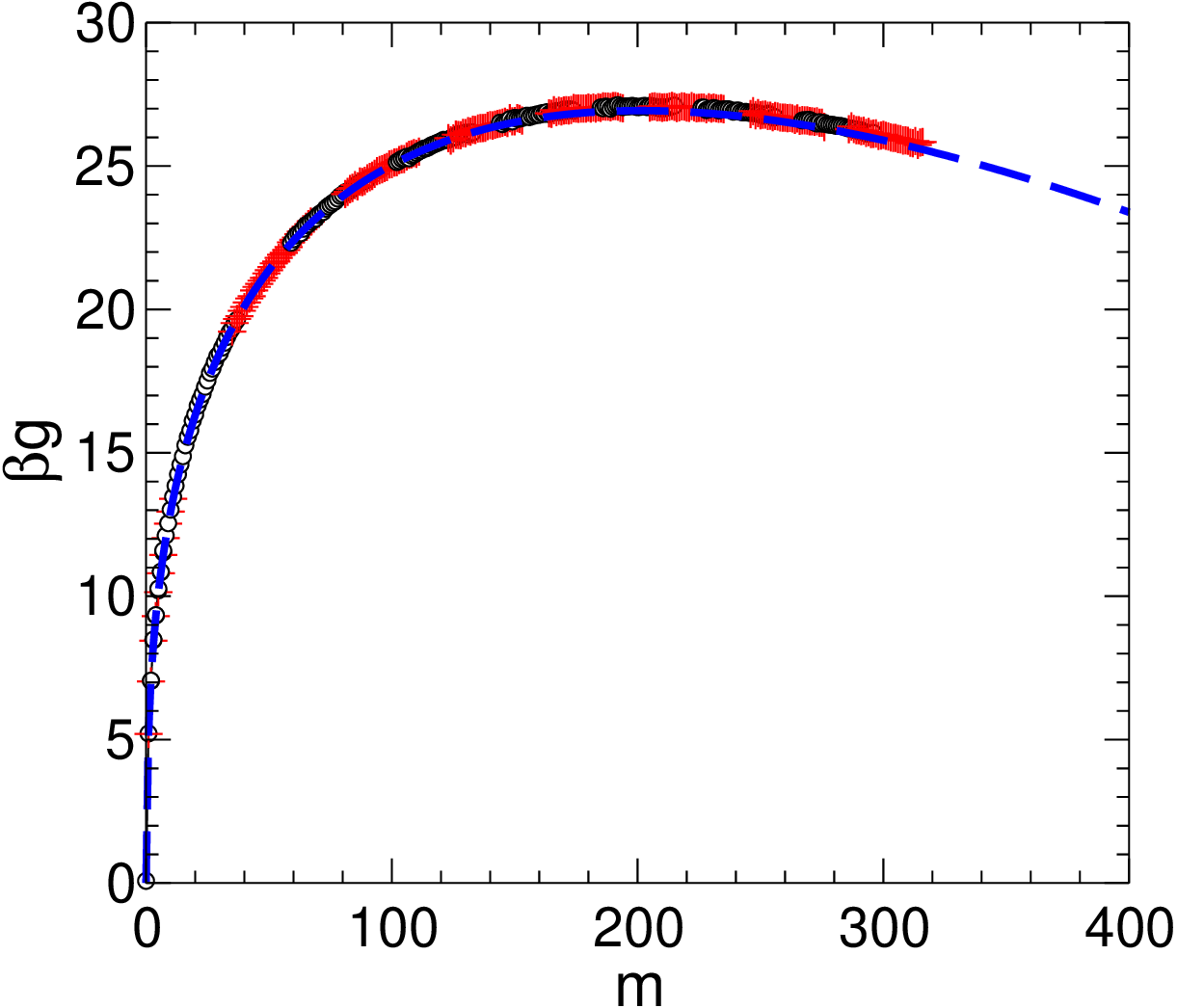}}
\caption{(Color online) Free energy barrier $g(m)$ for homogenous nucleation at $\beta J=0.65$ and $\beta H=0.05$.  The symbols (circles and plus signs) give the result obtained from umbrella sampling simulations using $\kappa/J=0.01$.  The full curve is constructed from 16 separate simulations, where $M_0$ is varied from 0 to 300 in steps of 20.  The data sets corresponding to each value of $M_0$ are represented by either black open circles or red plus signs.  
The dashed blue line gives our estimate for $g(m)$ obtained from CNT using the approach of Ryu and Cai~\cite{Cai2010a,Cai2010b}.
}
\label{us}
\end{figure}

By carrying out several simulations each for a different choice of $M_0$, we obtain estimates for overlapping segments of $g(m)$, which are then spliced together to form the complete $g(m)$ curve, as shown in Fig.~\ref{us}.  From the location of the maximum in $g(m)$, we find that the free energy barrier for homogeneous nucleation is approximately $27kT$, and that the size of the critical nucleus is approximately 200.  

By way of comparison, we note that Ref.~\cite{sear} estimates that the homogeneous nucleation barrier is $22kT$, and that the size of the critical nucleus is 219.  These estimates are obtained using classical nucleation theory (CNT), assume a square droplet, and use the Onsager result for the interfacial tension in the 2D Ising model.  Using a forward-flux sampling method, Ref.~\cite{sear} also estimates the homogeneous nucleation rate of the present system to be $3.3\times 10^{-13}$ events per MCS per lattice site.  Thus the homogeneous nucleation time for a system of $L^2=2025$ sites is $1.5\times 10^{9}$ MCS.

A more accurate procedure for evaluating $g(m)$ from CNT has recently been described by Ryu and Cai~\cite{Cai2010a,Cai2010b}.  Although the two key ingredients for CNT, the surface tension and the difference in chemical potential between the metastable and stable phases, are known exactly for the 2D Ising model, Ryu and Cai found that the standard CNT expression for $g(m)$ failed to fit simulation-based calculations of the free energy barrier.  However, by adding two additional terms to the CNT expression, one for shape fluctuations, and a constant term that ensures that the free energy of a single spin [i.e. $g(1)$] is correct, they were able to predict the free energy of forming a cluster within 1\% of their umbrella sampling simulation results, with no fitting parameters, over a wide of temperatures and field strengths. Our evaluation of $g(m)$ using Ryu and Cai's corrected CNT expression (Eq.~6 of Ref.~\cite{Cai2010a}) is included in Fig.~\ref{us} and shows a similar level of agreement with our simulation results.

As shown in the following sections, for the cases of heterogeneous nucleation studied here, the height of the heterogeneous nucleation barrier is always less than $14kT$, and the system nucleation time is always less than $5\times10^6$~MCS.  Heterogeneous nucleation processes are thus always more than 300 times faster than the homogeneous process, under all conditions studied here.  On this basis, we are assured that homogenous nucleation events (i.e. events that do not involve the impurity sites introduced below) are rare relative to heterogeneous events, and can be neglected in our analysis of the nucleation process in the presence of an impurity.

The above considerations also justify the choice of the system size ($L=45$) used here and in Ref.~\cite{sear}.  Since the homogeneous nucleation time of the system is proportional to $N$, then the smaller the system, the easier it is for heterogeneous nucleation events triggered by a single impurity to dominate the transformation of the metastable to the stable phase.  At the same  time, the system must be chosen large enough so that the critical cluster does not interact with its images across the periodic boundaries.  For both homogeneous and heterogeneous nucleation, we find that the size of the critical nucleus is always 200 or less.  In a system of size $N=45^2=2025$, the critical cluster will thus occupy 10\% or less of the total system volume.  Furthermore, since we conduct our simulations close to the coexistence curve at $H=0$, and well away from the critical temperature, we expect that the critical nucleus will be a relatively compact cluster, and that spin-spin correlations are negligible beyond a few lattice spacings.  It is thus extremely unlikely for a critical cluster in our system to interact with spins in its periodic images.

\begin{figure}[t]
\vskip 0.25in
\centerline{\includegraphics[scale=0.3]{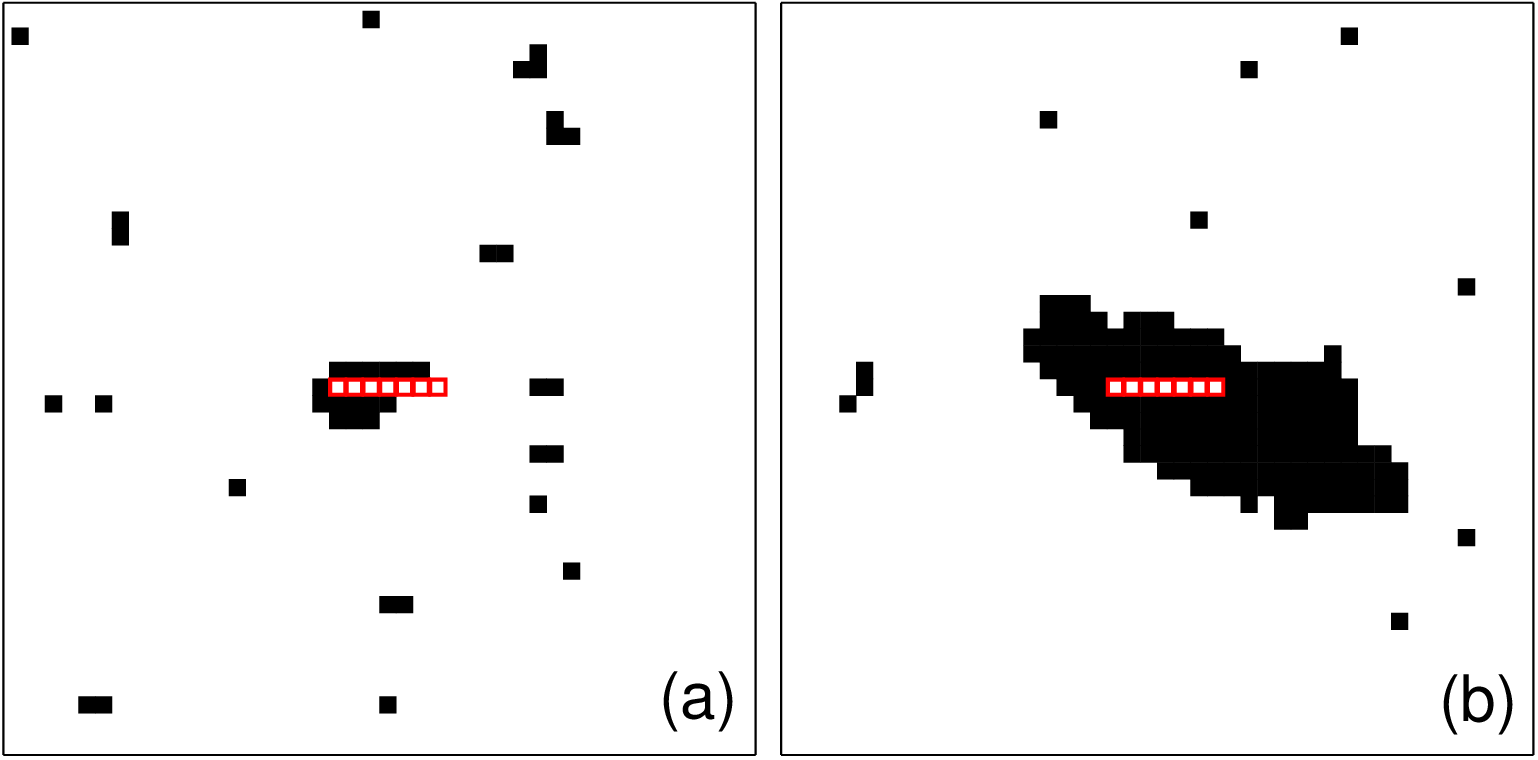}}
\caption{(Color online) Example configurations of the system in the metastable phase with $\beta J=0.65$, $\beta H=+0.05$, and $l=7$ (a) when $n=n_0=21$, and (b) when $n=n^\ast=174$.  Impurity sites are shown as red open squares.  Up-spins ($s_i=+1$) are shown as black filled squares.  White regions correspond to down-spins ($s_i=-1$).}
\label{pic}
\end{figure}

\section{Free energy barrier for Heterogeneous Nucleation}

To induce heterogeneous nucleation, we next study the case where our system contains an impurity consisting of a line of $l$ spins fixed to $s_i=+1$; see Fig.~\ref{pic}.  To find the free energy barrier for heterogeneous nucleation, we seek to evaluate the minimum reversible work of formation of a critical cluster of the stable phase.  However, since homogenous nucleation can be neglected, the critical cluster is necessarily a cluster of up-spins (i.e. sites with $s_i=+1$) attached to the impurity.  
In the following, we define the ``impurity cluster" as the contiguous cluster of up-spins that contains the impurity spins; thus the number of spins $n$ in the impurity cluster includes the impurity spins themselves.  Under this definition there can only be one impurity cluster, and so $n$ is a {\it system} property (and hence an order parameter) with respect to which the nucleation free energy barrier may be defined.  

To define the free energy barrier for heterogeneous nucleation, we first consider the partition function of the system for fixed $(N,H,T,l)$.  We write the system partition function ${\cal Z}=\sum_{n=l}^N Z(n)$ as a sum over the conditional partition function $Z(n)=\sum_{c(n)} \exp(-\beta E_c)$.  The sum in $Z(n)$ is over all system configurations $c$ in which the impurity cluster consists of exactly $n$ spins.  The corresponding conditional free energy is $G(n)=-kT\log Z(n)$, which is the free energy of the system when it contains an impurity cluster of size $n$.

\begin{figure}
\bigskip
\centerline{\includegraphics[scale=0.35]{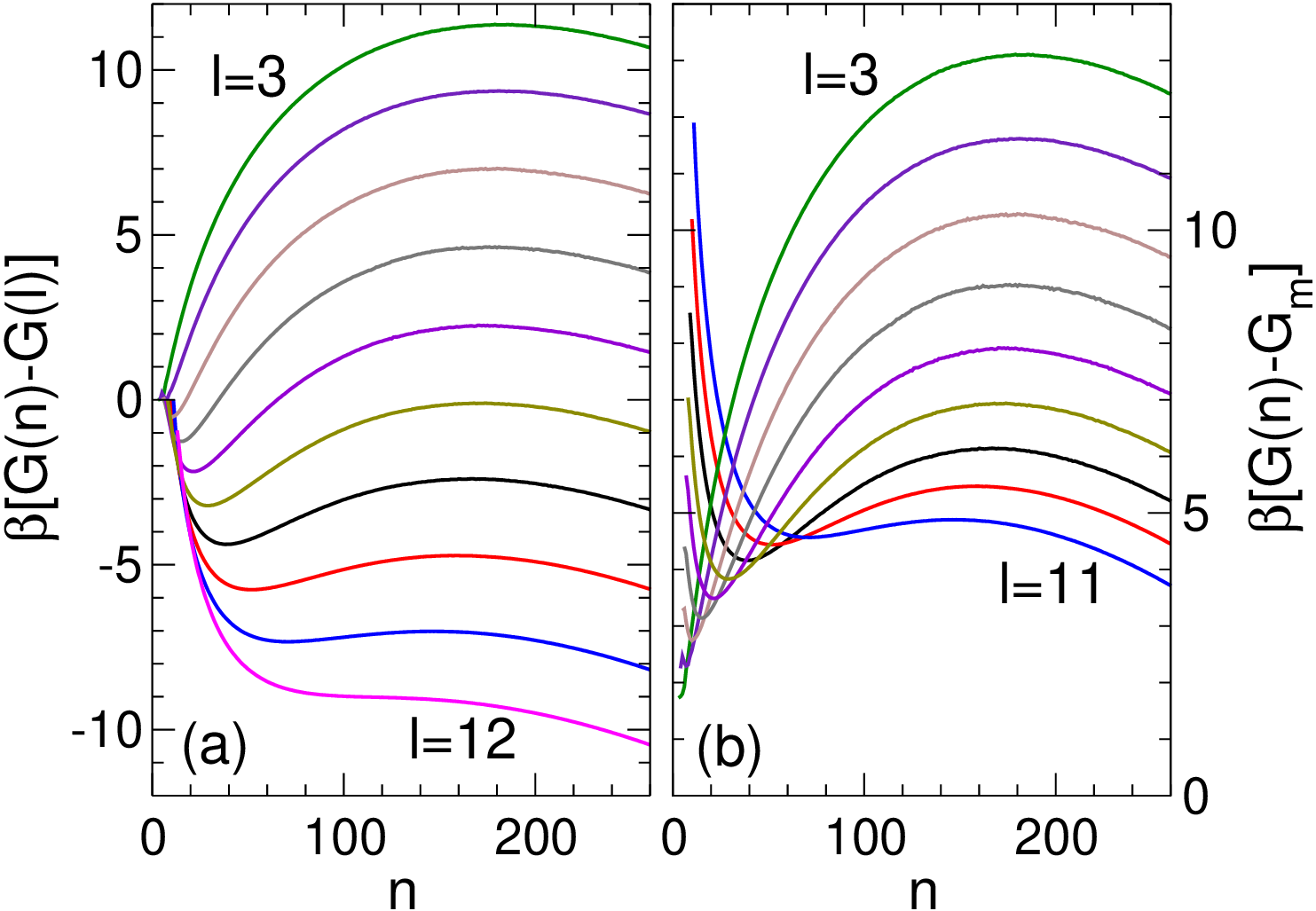}}
\caption{(Color online) (a) Free energy relative to a ``bare" impurity, $G(n)-G(l)$ as a function of $n$ for impurities of size $l=3$ to $12$.  (b) Free energy relative to the equilibrium metastable phase, $G(n)-G_m$ versus $n$ for $l=3$ to $11$.  For all curves, the statistical error is less than $0.02kT$.}
\label{barrier}
\end{figure}

To compute $G(n)$, we note that the probability to observe an impurity cluster of $n$ spins is $P(n)=Z(n)/{\cal Z}$. Consequently, the work of formation of an $n$-spin impurity cluster, starting from a ``bare" impurity (i.e. $n=l$), is given by the free energy difference,
\begin{eqnarray}
G(n)-G(l) = -kT\log \frac{Z(n)}{Z(l)}
= -kT\log \frac{P(n)}{P(l)}.
\label{hetg}
\end{eqnarray}

We evaluate $G(n)$ using Eq.~\ref{hetg} from simulations in which $l$ ranges from 3 to 12.  As shown below, for this range of $l$ the variation of $G(n)$ is never more that $12kT$, and therefore multi-window umbrella sampling is not required.  Rather, we simply impose a constraint on our MC sampling such that $n\leq n_{\rm max}=300$.   This choice of $n_{\rm max}$ restricts our simulations to the metastable phase, and to configurations in the vicinity of transition states to the stable phase.  This approach is equivalent to using a single umbrella sampling window in which $U_B=0$ for $n\leq n_{\rm max}$ and $U_B=\infty$ for $n>n_{\rm max}$.  As in all umbrella sampling simulations, the relative probabilities with which configurations occur inside the umbrella window are correctly estimated after the appropriate reweighting, regardless of the specific form of $U_B$.   Consequently, our results for $n\leq n_{\rm max}$ are independent of the choice of $n_{\rm max}$.

We evaluate the equilibrium ratio $P(n)/P(l)$ from our simulations, and plot the result for $G(n)-G(l)$ in Fig.~\ref{barrier}(a), for various $l$.  For $3\leq l \leq 11$, each curve exhibits a maximum at $n=n^\ast$, indicating the size of the critical cluster.  The value of $n^\ast$ demarcates the boundary between the metastable and stable phases of the system, and we define the configuration space of the metastable phase as the set of microstates for which $l\leq n\leq n^\ast$.  

\begin{figure}
\centerline{\includegraphics[scale=0.34]{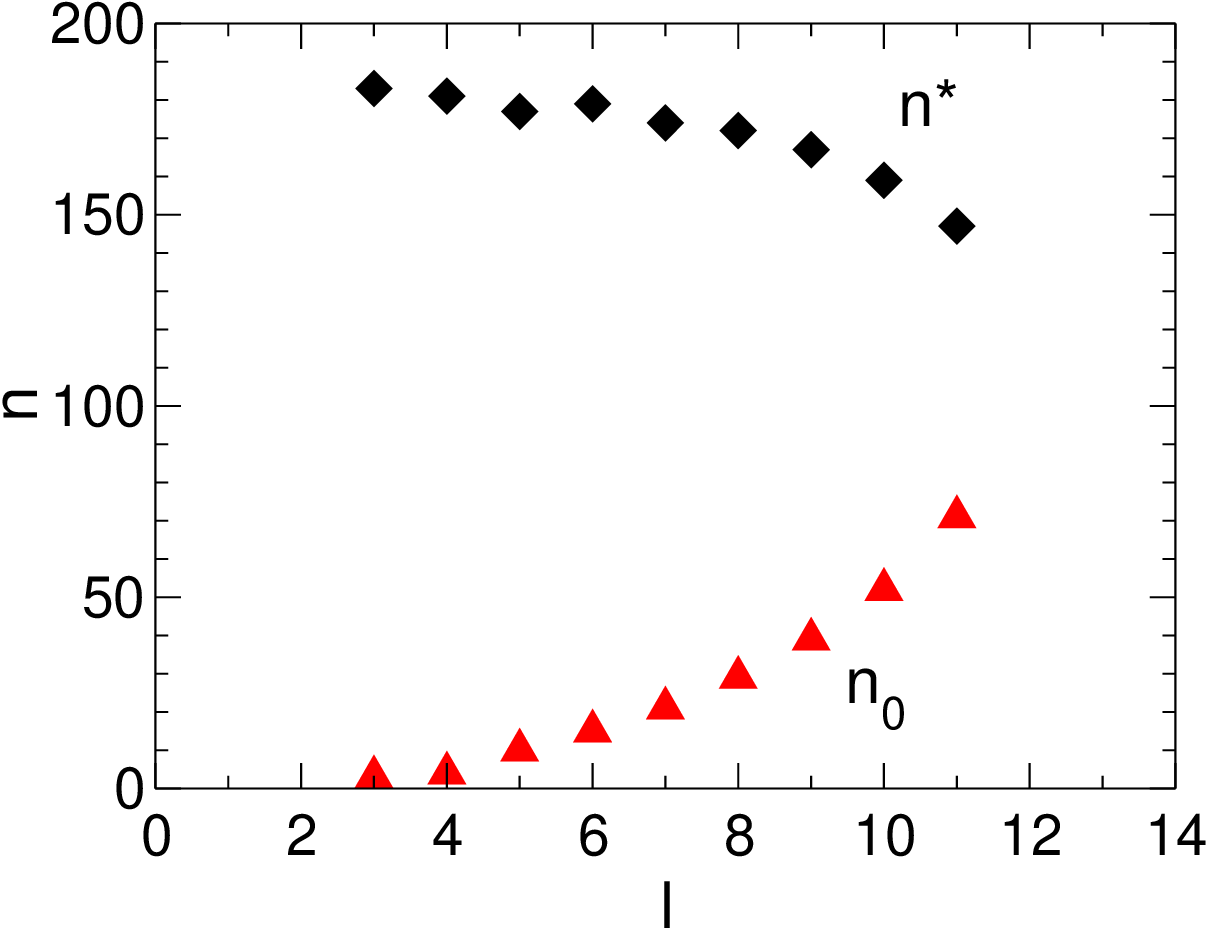}}
\caption{(Color online) Size of the impurity cluster at the maximum ($n^\ast$) and minimum ($n_0$) of $G(n)$ as a function of impurity size $l$.  Statistical errors are smaller than the symbol size.}
\label{n}
\end{figure}

Fig.~\ref{barrier}(a) also shows that as $l$ increases, a minimum in $G(n)$ at $n=n_0$ emerges and grows; this feature corresponds to wetting of the impurity by a finite cluster of the stable phase.  
For $l=12$, $G(n)$ is a monotonically decreasing function of $n$, and the metastable phase has ceased to exist.  This qualitative change in the shape of $G(n)$ as $l$ increases thus represents the limit of stability of the metastable phase.  This limit of stability is also seen in Fig.~\ref{n}, where we show that $n^\ast$ and $n_0$ approach one another, and then become undefined for $l\geq 12$.  

To obtain the free energy barrier for nucleation from $G(n)$, we must take care to identify the appropriate thermodynamic reference state with respect to which the barrier height should be measured.  
We follow the reasoning of Ref.~\cite{frenkel-true}, which studied homogeneous nucleation, adapted here for the case of heterogeneous nucleation.  That is, the free energy barrier for nucleation is defined as the minimum reversible work required to apply a constraint that confines the system to the transition state at $n=n^\ast$, starting from a reference state that considers the entire configuration space of the metastable phase, i.e. all configurations in the range $l\leq n\leq n^\ast$.  
To implement this definition, we define the partition function of the metastable phase ${\cal Z}_m=\sum_{n=l}^{n^\ast} Z(n)$ as a restricted sum over all states such that $l\leq n\leq n^\ast$.
The corresponding free energy of the metastable phase is $G_m=-kT\log {\cal Z}_m$.  The work of formation of an $n$-spin impurity cluster, starting from the equilibrium metastable phase, is then given by the free energy difference,
\begin{eqnarray}
G(n)-G_m &=& -kT\log \frac{Z(n)}{{\cal Z}_m} \cr
&=& -kT\log \frac{P(n)}{\sum_{n'=l}^{n^\ast} P(n')}.
\label{eq:qm}
\end{eqnarray}
The second equality above emphasizes that $G(n)-G_m$ can also be evaluated in our simulations from the relative probabilities for observing the impurity cluster to have various $n$.  

\begin{figure}
\centerline{\includegraphics[scale=0.35]{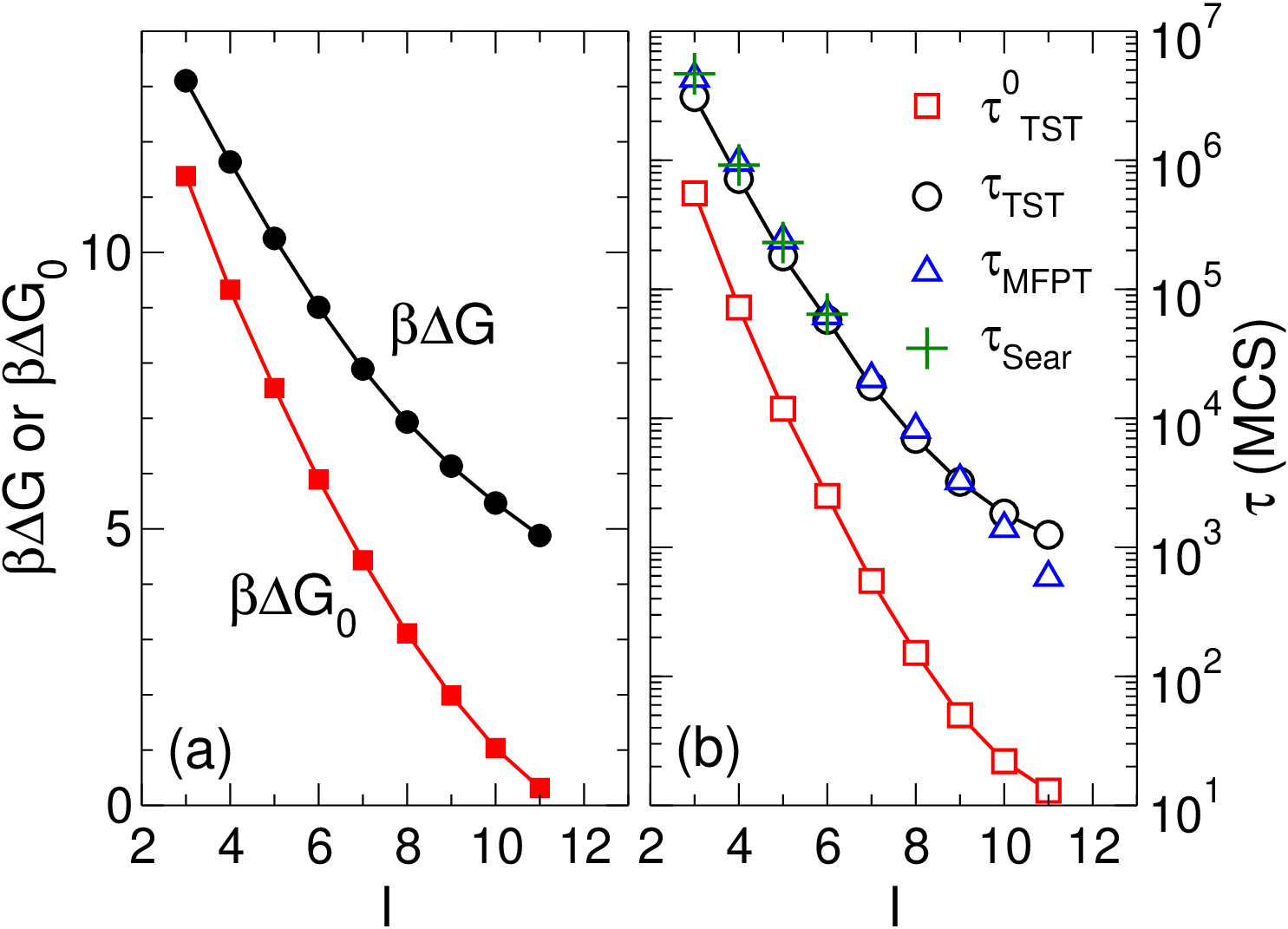}}
\caption{(Color online) (a) Comparison of the nucleation barriers $\Delta G$ and $\Delta G_0$ as a function of impurity size $l$. (b) Nucleation times $\tau^0_{\rm TST}$, $\tau_{\rm TST}$, $\tau_{\rm MFPT}$, and $\tau_{\rm Sear}$ as a function of $l$.  For comparison, note that the homogeneous nucleation time when no impurity is present is $1.5\times10^9$~MCS~\cite{sear}.  For all quantities, statistical errors are smaller than the symbol size.}
\label{G}
\end{figure}

Our results for $G(n)-G_m$ are plotted in Fig.~\ref{barrier}(b).  
The difference between the free energy curves in Fig.~\ref{barrier}(a) and (b) is a change in the reference state, giving rise to an $l$-dependent 
vertical shift without a change in shape.
The work of formation of the transition state from the metastable phase (i.e. the free energy barrier for nucleation) is given by $\Delta G = G(n^\ast)-G_m$.  
As shown in Fig.~\ref{G}(a), $\Delta G$ does not go to zero at the limit of stability.  Although paradoxical at first glance, this result is physically reasonable for our system.  Since $n^\ast$ remains non-zero even at the limit of stability, the metastable phase encompasses a considerable region of configuration space ($l\leq n\leq n^\ast$) up to the point where stability is lost.  Hence the work required to create the transition state remains finite, even as the metastable state ceases to exist as a distinct phase.  In previous work, it has been assumed that the nucleation barrier should go to zero as the thermodynamic stability of a metastable phase is lost~\cite{BZ,TO,parrinello06,vanish,S}.  Our system provides a counter-example.

\section{Nucleation time}

We next assess the implications of our results for $\Delta G$ for estimating the nucleation time using TST~\cite{Cai2010a,Cai2010b}.  For our system, the TST prediction for the nucleation time is,
\begin{equation}
\tau_{\rm TST}=(f^+_c z)^{-1}\exp(\beta \Delta G),
\label{TST}
\end{equation}
where $\tau_{\rm TST}$ is the average time (in MCS) per impurity for a critical cluster to appear in the system that subsequently evolves into the stable phase.
$z=\sqrt{\beta \eta/2 \pi}$ is the Zeldovich factor, where $\eta=-(\partial^2 G/\partial n^2)_{n=n^\ast}$ is the curvature of $G(n)$ at the top of the barrier.  We estimate $\eta$ from a quadratic fit to data that lies within $0.2kT$ of the maximum of $G(n)$.  $f^+_c$ is the attachment rate of monomers to the critical cluster.  We determine $f^+_c$ from the time dependence of fluctuations of the size of critical clusters, following the same procedure used in Refs.~\cite{Cai2010a,Cai2010b}.  The result for $\tau_{\rm TST}$ obtained from our data is shown in Fig.~\ref{G}(b).

\begin{figure}
\centerline{\includegraphics[scale=0.34]{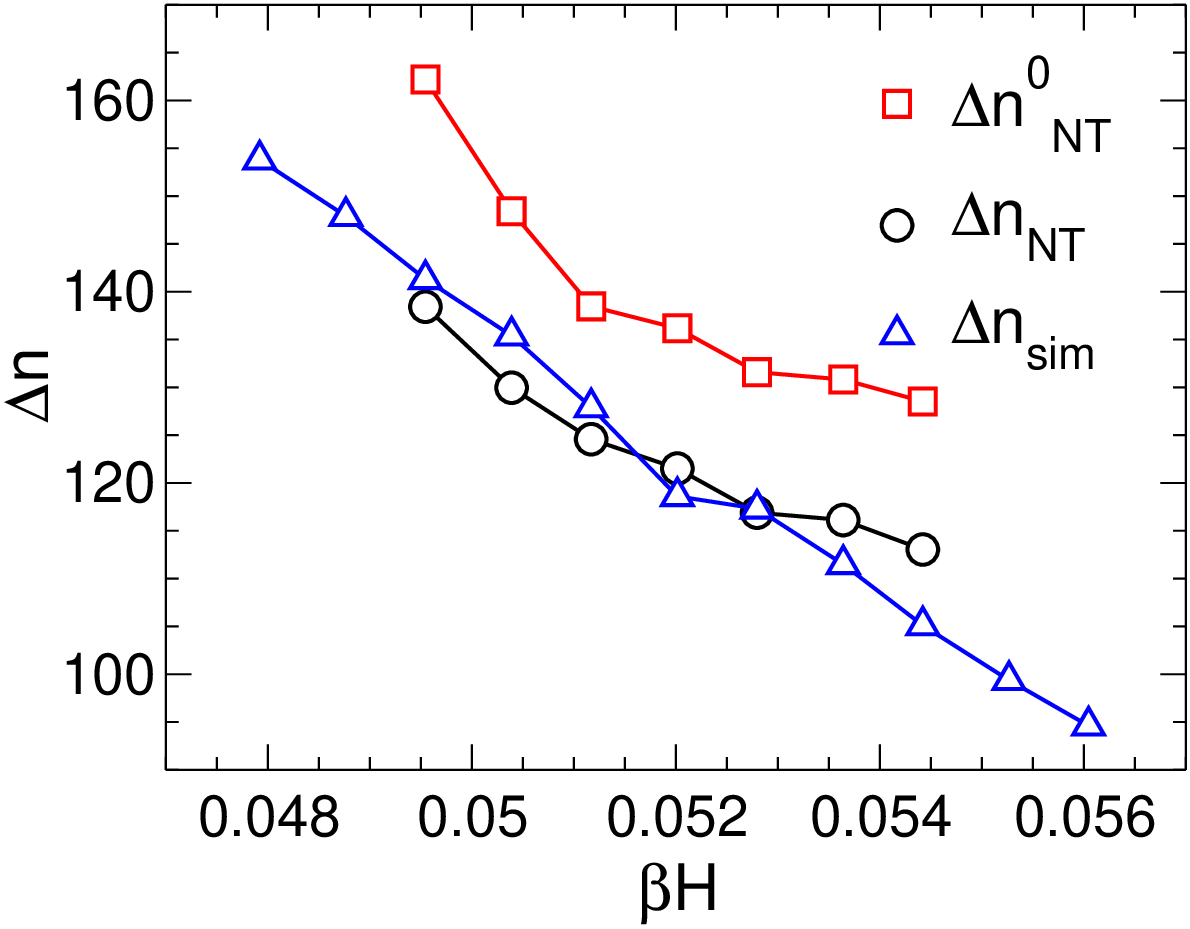}}
\caption{(Color online) Comparison of our estimates for the excess number of up-spins in the critical cluster $\Delta n_{\rm NT}$, $\Delta n^0_{\rm NT}$, and $\Delta n_{\rm sim}$ as a function of $\beta H$, for the $l=7$ system.}
\label{thm}
\end{figure}

To test the accuracy of $\tau_{\rm TST}$, we directly evaluate the nucleation time in terms of the mean first passage time (MFPT) for the impurity cluster to grow to the critical size.  For a given $l$, we set $n_{\rm max}=n^\ast$ so that the system is confined to explore only the configuration space of the metastable phase, and bring this constrained system into equilibrium.  Then, at a randomly selected time, we set $t=0$ and measure the time it takes for the system to first reach $n=n_{\rm max}$.  The MFPT is the average of many such measurements.  We define the nucleation time $\tau_{\rm MFPT}$ as twice the MFPT, because only half of the runs that reach the transition state would ultimately evolve into the stable phase. As shown in Fig.~\ref{G}(b), $\tau_{\rm TST}$ is in excellent agreement with $\tau_{\rm MFPT}$.  Fig.~\ref{G}(b) also shows that $\tau_{\rm MFPT}$ is consistent with the nucleation times ($\tau_{\rm Sear}$) reported in Ref.~\cite{sear} for the same system, as found using a forward-flux sampling method.  Our results thus demonstrate that in our case TST is capable of predicting the nucleation time with remarkable accuracy even at the very limit of stability of the metastable phase.

\section{Nucleation Theorem}

It is also possible to validate our results for $\Delta G$ by testing the nucleation theorem.  The nucleation theorem~\cite{kash1982,oxtoby1994,ford1996,bowles2001,kash2006} states that,
\begin{equation}
\left(\frac{\partial \Delta G}{\partial \Delta \mu}\right)_{T}=\frac{1}{2}\left(\frac{\partial \Delta G}{\partial H}\right)_{T}=-\Delta n\mbox{,}\\
\label{NT}
\end{equation}
where $\Delta \mu$ is the difference in chemical potential between the stable and metastable phases, and $\Delta n$ is the excess number of up-spins in the critical cluster.  In the first equality, we have used $\Delta \mu\approx2H$, which for the Ising model is a good approximation for $T$ below the Curie temperature~\cite{Cai2010a,Cai2010b,harris1984}. To conduct this test, we carry out new runs for the case of $l=7$ over a range of $\beta H$ from $0.048$ to $0.056$.  Although $\Delta n$ can be approximated as $n^\ast - n_0$, in the low barrier regime it is more accurate to directly evaluate $\Delta n$ as the difference in the average number of up-spins in the entire system (including those not in the impurity cluster) when the system is at $n=n^\ast$, and the average number of up-spins in the metastable phase averaged over all $l\leq n\leq n^\ast$; these results are shown in Fig.~\ref{thm} and denoted as $\Delta n_{\rm sim}$.  We also evaluate $\Delta G$ as a function of $H$, and estimate the derivative in Eq.~\ref{NT} using a five-point central-difference numerical method.  The estimate of $\Delta n$ thus obtained from Eq.~\ref{NT} is denoted $\Delta n_{\rm NT}$ in Fig.~\ref{thm}, and is in good agreement with $\Delta n_{\rm sim}$.

\section{Comparison of barrier definitions}

Although our definition of $\Delta G$ is straightforward, we note that almost all previous studies of heterogeneous nucleation on small impurities use a different definition.  Specifically, when the free energy as a function of $n$ exhibits both a minimum (at $n_0$) and a maximum (at $n^\ast$)  the nucleation barrier is usually defined as $\Delta G_0=G(n^\ast)-G(n_0)$~\cite{cacciuto05,vanish,kelton,CK,OR,BZ,TO,WWY,KO,AF,S,S07,DD,oh01}.  However, this definition is an approximation that becomes increasingly inaccurate in the low barrier regime.  
To illustrate the problem, we show our results for $\Delta G_0$ as a function of $l$ in Fig.~\ref{G}(a).  Whereas $\Delta G$ remains finite at the limit of stability, $\Delta G_0$ vanishes.
In Fig.~\ref{G}(b), we show $\tau^0_{\rm TST}$, the TST prediction for the nucleation time obtained if we use $\Delta G_0$ instead of $\Delta G$ in Eq.~\ref{TST}.  We find that for the lowest barriers (at large $l$), $\tau^0_{\rm TST}$ underestimates $\tau_{\rm MFPT}$ by more that two orders of magnitude.
Similarly, if we use $\Delta G_0$ in Eq.~\ref{NT}, the estimate obtained for $\Delta n$ (denoted $\Delta n^0_{\rm NT}$) is distinctly less accurate than that found using $\Delta G$ (Fig.~\ref{thm}).  

The above results demonstrate that the use of $\Delta G_0$ instead of $\Delta G$ leads to a qualitatively different and erroneous physical picture for nucleation in the low barrier regime:  Using $\Delta G_0$, the barrier vanishes, and theories such as TST break down, whereas using $\Delta G$ we find that the actual behavior is exactly the opposite.
We emphasize that the difference between $\Delta G$ and $\Delta G_0$ is only apparent in the low barrier regime.  When the barrier is high, even small clusters are rare, and the properties of the metastable phase are dominated by system configurations found near $n=l$.  In this limit $G(l)\approx G_m$, and $\Delta G$ and $\Delta G_0$ become equivalent.  However, when approaching a limit of stability, the correct definition of the free energy barrier must be used.

\section{Discussion}

It is important to note how the definition of the free energy of cluster formation for the homogeneous system [$g(m)$ in Eq.~\ref{homo}] differs from that for the heterogeneous case [$G(n)$ in Eq.~\ref{hetg}] in the low barrier regime.
The definition of $g(m)$ is correct in the limit that stable-phase clusters are rare and non-interacting.  In a finite-sized system near the transition state, this limit is realized only if there is at most one large cluster in the system.
However, when the homogeneous nucleation barrier approaches $kT$, several large clusters may form simultaneously.   In this case, cluster interactions cannot be neglected, and Eq.~\ref{homo} is no longer accurate.
In contrast, our definition of $G(n)$ for heterogeneous nucleation depends only on taking the limit that homogeneous nucleation events are rare, which we have assured by our choice of $T$, $H$, and $N$.
By construction, there is always one, but only one, impurity cluster of any size present in our heterogeneous system, regardless of the height of the heterogeneous nucleation barrier.  As a consequence, multiple large clusters do not occur in our system, even as we approach the stability limit of the metastable phase, and thus Eq.~\ref{hetg} does not break down when the barrier height approaches $kT$. 

We also emphasize that our definitions of the metastable phase and its limit of stability are only well-defined for a finite-sized system.   As discussed in Section III, our system size is deliberately chosen to be small enough so that homogeneous nucleation processes can be neglected.  If we take the limit $N\to \infty$ in a system that contains only one impurity, a homogeneous nucleation event somewhere in the system becomes overwhelmingly more probable than an event triggered by a lone impurity.  This is a well-known conceptual challenge associated with the definition of metastability for any system (homogeneous or heterogeneous) in the thermodynamic limit~\cite{pablo,langer,penrose}.

In addition, our results demonstrate the importance of the reference state when calculating the nucleation rate from a measure of the nucleation barrier.  This insight is facilitated here by the fact that the definition of the heterogeneous nucleation barrier is free of the complications that arise in the homogenous case when the barrier is low, as discussed above.  Although further work is required, we anticipate that a similar examination of the reference state appropriate to homogeneous nucleation may elucidate the rate and its relation to thermodynamic quantities in the low barrier regime.

In summary, for heterogeneous nucleation on small impurities, we show that $\Delta G$ remains well-defined, and does not vanish, at the limit of stability of the metastable phase.  Furthermore, we find that both TST and the nucleation theorem are impressively accurate, even at the limit of stability, so long as the correct reference state is used to define the height of the nucleation barrier.  
We expect that the pattern of behavior found here will be common to all low-barrier systems where a free energy minimum and maximum converge at a finite value of the order parameter, and thus may be generic for heterogeneous nucleation on small impurities.  That is, for impurity-induced nucleation, it is only when the size of the critical cluster goes to zero ($n^\ast \to 0$) that we should expect the nucleation barrier to vanish at the limit of stability.

\section{acknowledgements} 
We thank ACEnet and WestGrid for providing computational resources, and NSERC for financial support.  PHP thanks the CRC program for support.

\end{document}